\documentclass[english]{IEEEtran}
\usepackage[LGR,T1]{fontenc}
\usepackage[latin9]{inputenc}
\usepackage{babel}
\usepackage{array}
\usepackage{float}
\usepackage{booktabs}
\usepackage{multirow}
\usepackage{amstext}
\usepackage{graphicx}
\usepackage[unicode=true]
 {hyperref}

\makeatletter

\DeclareRobustCommand{\greektext}{%
  \fontencoding{LGR}\selectfont\def\encodingdefault{LGR}}
\DeclareRobustCommand{\textgreek}[1]{\leavevmode{\greektext #1}}
\ProvideTextCommand{\~}{LGR}[1]{\char126#1}

\providecommand{\tabularnewline}{\\}
\floatstyle{ruled}
\newfloat{algorithm}{tbp}{loa}
\providecommand{\algorithmname}{Algorithm}
\floatname{algorithm}{\protect\algorithmname}

\makeatother

\begin{document}
\title{Efficient Fault Detection Architectures for Modular Exponentiation
Targeting Cryptographic Applications Benchmarked on FPGAs}
\author{Saeed Aghapour\textit{, }Kasra Ahmadi, Mehran Mozaffari Kermani, \textit{Senior
Member}, \textit{IEEE, }and Reza Azarderakhsh\textit{, Member, IEEE}}
\IEEEspecialpapernotice{\thanks{S. Aghapour, K. Ahmadi, and M. Mozaffari-Kermani are with the Department
of Computer Science and Engineering, University of South Florida,
Tampa, FL 33620, USA. e-mails: \{aghapour, ahmadi1, mehran2\}@usf.edu.} \thanks{R. Azarderakhsh is with the Department of Computer and Electrical
Engineering and Computer Science, Florida Atlantic University, Boca
Raton, FL 33431, USA. e-mails: \{razarderakhsh\}@fau.edu.}}
\maketitle
\begin{abstract}
Whether stemming from malicious intent or natural occurrences, faults
and errors can significantly undermine the reliability of any architecture.
In response to this challenge, fault detection assumes a pivotal role
in ensuring the secure deployment of cryptosystems. Even when a cryptosystem
boasts mathematical security, its practical implementation may remain
susceptible to exploitation through side-channel attacks. In this
paper, we propose a lightweight fault detection architecture tailored
for modular exponentiation\textemdash a building block of numerous
cryptographic applications spanning from classical cryptography to
post quantum cryptography. Based on our simulation and implementation
results on ARM Cortex-A72 processor, and AMD/Xilinx Zynq Ultrascale+,
and Artix-7 FPGAs, our approach achieves an error detection rate close
to 100\%, all while introducing a modest computational overhead of
approximately 7\% and area overhead of less than 1\% compared to the
unprotected architecture. To the best of our knowledge, such an approach
benchmarked on ARM processor and FPGA has not been proposed and assessed
to date.
\end{abstract}


\section{Introduction}

In today's era of online communication, cryptography plays an essential
role for secure interaction. However, besides pure mathematical analysis,
cryptographic algorithms can be threatened by exploitation of their
implementation, known as side-channel attacks. Numerous studies stress
the importance of enhancing the side-channel security of cryptographic
algorithms \cite{side channel 1,Side 2,side 3}, and \cite{new dr}.

One type of side-channel attacks is known as fault analysis attack
which was introduced in \cite{bone 1} and \cite{fault- start}. In
fault attacks, adversaries intentionally induce malfunctions in a
cryptosystem, hoping that these faults will reveal secret values within
the system. These malfunctions can be caused by injecting faulty inputs
into the algorithm or by disrupting its normal functionality. This
poses a significant concern in cryptography, where even a single bit
change can lead to entirely different outputs. The research work of
\cite{fault inject 1} provides a comprehensive study on different
fault injection methods that do not require expensive equipment. 

As a countermeasure to fault attacks, fault detection schemes have
been developed. Various fault detection techniques have been introduced
for different components of both classical and post-quantum cryptosystems,
spanning both symmetric or asymmetric cryptography. For instance,
\cite{AES 1} and \cite{AES 2} presented efficient fault detection
schemes for the AES with very high error coverage, and, \cite{RSA 1}
and \cite{RSA 2} proposed fault detection schemes for RSA.

A number of research works have been presented to address fault detection
in elliptic curve cryptography, with a particular focus on the scalar
multiplication (ECSM) component. In \cite{DR Hassan}, a novel fault
detection scheme based on recomputation for ECSM is presented, offering
extensive error coverage while imposing minimal computational overhead.
Additionally, the works in \cite{Kasra window} and \cite{Koblitz}
have proposed highly efficient fault detection methods for both ECSM
and \textgreek{t}NAF conversion, capable of detecting transient and
permanent errors with success rate of close to 100\%. Moreover, there
have been efforts to propose a generic approach leveraging deep learning
for detecting vulnerabilities in ciphers against fault attacks, as
outlined in \cite{deep learning}.

As technology advances and we approach the advent of practical quantum
computers, Shor's algorithm \cite{Shor} underscores the urgency of
shifting from classical cryptography to new standard Post-Quantum
Cryptography (PQC) schemes. Consequently, significant research efforts
have been directed toward proposing fault detection mechanisms for
the new standard schemes, with a particular emphasis. 

In \cite{pqc 1}, an assessment of existing countermeasures and their
associated computational overheads for protecting lattice-based signature
schemes against fault attacks is presented. Moreover, Sarker et al.
\cite{pqc 2} proposed an error detection algorithm for number theoretic
transform (NTT), which could be deployed on any lattice based scheme
that uses this operation. Additionally, the work in \cite{pqc 4 iscas}
proposes fault attack countermeasures for error samplers which are
employed within lattice-based schemes to introduce noise to the secret
information, thereby concealing direct computations involving that
sensitive data.

\textbf{Our motivations:} Having discussed the previous research,
limited attention has been given to fault detection solely for the
exponentiation module. The significance of such work lies in its applicability
across a wide range of applications employing this module and not
being limited to a single cryptosystem. For instance, KAZ \cite{KAZ}
which is a PQC candidate scheme in the July 2023 new NIST's additional
signature competition, relies extensively on modular exponentiation.

To the best of our knowledge, the first work proposing a generic fault-resistant
method for exponentiation is \cite{fault-expo 1}. Their method imposes
a computational overhead of up to 50\% to the algorithm. However,
as demonstrated in \cite{break fumaroli}, the proposed method is
found to be insecure against fault attacks when applied to RSA-CRT
(RSA using Chinese Remainder Theorem). Furthermore, the work in \cite{fault-expo 2}
presented a new approach which besides resisting fault attacks, could
also resist against power analysis attacks. However, their work lacks
implementation details to provide practical insight.

In this paper, we aim to present a novel fault detection scheme tailored
for modular exponentiation, a fundamental component in numerous cryptosystems.
To the best of our knowledge, this is the first approach using partial
recomputation for such architectures, leading to low overhead on ARM
processors and FPGAs.

\section{Preliminaries}

One of the most efficient techniques for computing modular exponentiation
is called Right-to-Left algorithm \cite{Right to left}. To calculate
the value of $x^{y}\text{ mod }N$ using this approach, first $y$
is represented in binary form as $y=\sum_{i=0}^{n-1}a_{i}2^{i}$.
Therefore, $x^{y}\text{ mod }N$ can be expressed as $\prod_{i=0}^{n-1}x^{a_{i}2^{i}}\text{ mod}\,N$.
Now, starting from $i=0$, if $a_{i}=0$, the base is squared, and
we proceed to the next bit. However, if $a_{i}=1$, then the intermediate
result must be multiplied by $x$ before squaring the base. Algorithm
1 presents this approach.

\begin{algorithm}[t]
\textbf{~}

\textbf{Input: }base $x$, exponent $y$, and modulus $N$

\textbf{Output: $result=x^{y}$ }mod\textbf{ $N$}

1:~~$result=1$

2:~~$x=x$ mod $N$

3:~~\textbf{while} $(y>0)$

4:~~~~~~\textbf{if} $(y$\textbf{ }mod $2==1)$

5:~~~~~~~~~~$result=(result\times x)$ mod $N$

6:~~~~~~$y=y>>1$

7:~~~~~~$x=(x\times x)$ mod $N$

8:~~\textbf{return} $result$

~

\caption{Right-to-Left Exponentiation Algorithm}
\end{algorithm}

\section{Proposed Fault Detection Architecture}

In this section, we present our approach to detecting faults in the
modular exponentiation operation of $x^{y}\text{ mod }N$. Our method
relies on recomputation, where we perform extra calculations, and
the output is considered valid only if the results of the two calculations
match.

\subsection{Encoding/Decoding}

Encoding the inputs plays a crucial role in recomputation-based schemes.
This is because without encoding, permanent faults and identical transient
errors cannot be detected, as they would produce identical outputs
in both computations. An encoding module is tasked with generating
distinct input values for the two computations at times $t_{1}$ and
$t_{2}$. After these separate computations are performed on these
distinct inputs, a decoding algorithm should yield the same result
at both $t_{1}$ and $t_{2}$. Efficient encoding and decoding algorithms
should not introduce excessive computational overhead into the scheme.

\subsubsection{Encoding the Base}

In base encoding, we leverage the property of modular exponentiation,
which states that $x^{y}$ mod $N\equiv((x$ mod $N)^{y})$ mod $N$.
As a result, we select a random number $k_{x}$ and compute $x_{enc}=x+k_{x}N$.
Since $x_{enc}^{y}$ mod $N$ is equal to $x^{y}$ mod $N$, there
is no requirement for a decoding step in this encoding scheme.

\subsubsection{Encoding the Exponent}

To encode the exponent, we utilize a property from group theory, which
states that $x^{k.ord(x)}$ mod $N\equiv1$ where $ord(x)$ is the
order of the element $x$ in the group $G$ over modulus $N$. However,
calculating the order of $x$ for each $x$ in an algorithm could
be problematic. Therefore, we substitute $ord(x)$ with a multiple
of $\phi(N)$, where $\phi(N)$ is Euler's totient function. Similarly,
as $x^{k.\phi(N)}$ mod $N\equiv1$, we encode the exponent as $y_{enc}=y+k_{y}\phi(N)$.
Consequently, because $x^{y_{enc}}$ mod $N\equiv x^{y}$ mod $N$,
there is no need for a decoding step.

\subsection{Proposed Schemes}

In this subsection, we introduce two fault detection schemes. The
first scheme involves a full recomputation, while the second scheme
employs a partial recomputation approach to mitigate computational
overhead.

\begin{algorithm}[t]
\textbf{~}

\textbf{Input: }base $x$, exponent $y$, modulus $N$, $\phi(N)$,
and $l$

\textbf{Output: $result$}, $result_{partial}$, and $HM$

01:~~$result=1$

02:~~$x=x$ mod $N$

03:~~$y=y$ mod $\phi(N)$

04:~~$counter=0$

05:~~$HM=0$

06:~~\textbf{while} $(y>0)$

07:~~~~~~\textbf{if} $(y$\textbf{ }mod $2==1)$

08:~~~~~~~~~~$result=(result\times x)$ mod $N$

09:~~~~~~~~~~$HM++$

10:~~~~~~$y=y>>1$

11:~~~~~~$x=(x\times x)$ mod $N$

12:~~~~~~$counter++$

13:~~~~~~\textbf{if} $(counter==l)$

14:~~~~~~~~~~$result_{partial}=result$

15:~~\textbf{return} $result$, $result_{partial}$, $HM$

~

\caption{Our modular exponentiation module}
\end{algorithm}

\subsubsection{Scheme 1: Full Recomputation}

In this scheme the output is computed twice, compared, and accepted
only if the two outputs match. With more details, at time $t_{1}$,
after encoding the base and exponent as $x_{1}=x+k_{1}N$ and $y_{1}=y+k_{2}\phi(N)$,
the output $Q_{1}\equiv x_{1}^{y_{1}}\text{ mod }N$ is computed.
Similarly, for the recomputation at $t_{2}$, the base and exponent
are encoded as $x_{2}=x+k_{3}N$ and $y_{2}=y+k_{4}\phi(N)$, and
the output $Q_{2}\equiv x_{2}^{y_{2}}\text{ mod }N$ is computed.
The output is accepted only if $Q_{1}=Q_{2}$. This full recomputation
approach guarantees very close to 100\% error coverage but comes at
the cost of doubling the entire computation.

\begin{figure*}[t]
\centering

\includegraphics[bb=0bp 245bp 960bp 455bp,clip,width=1.8\columnwidth]{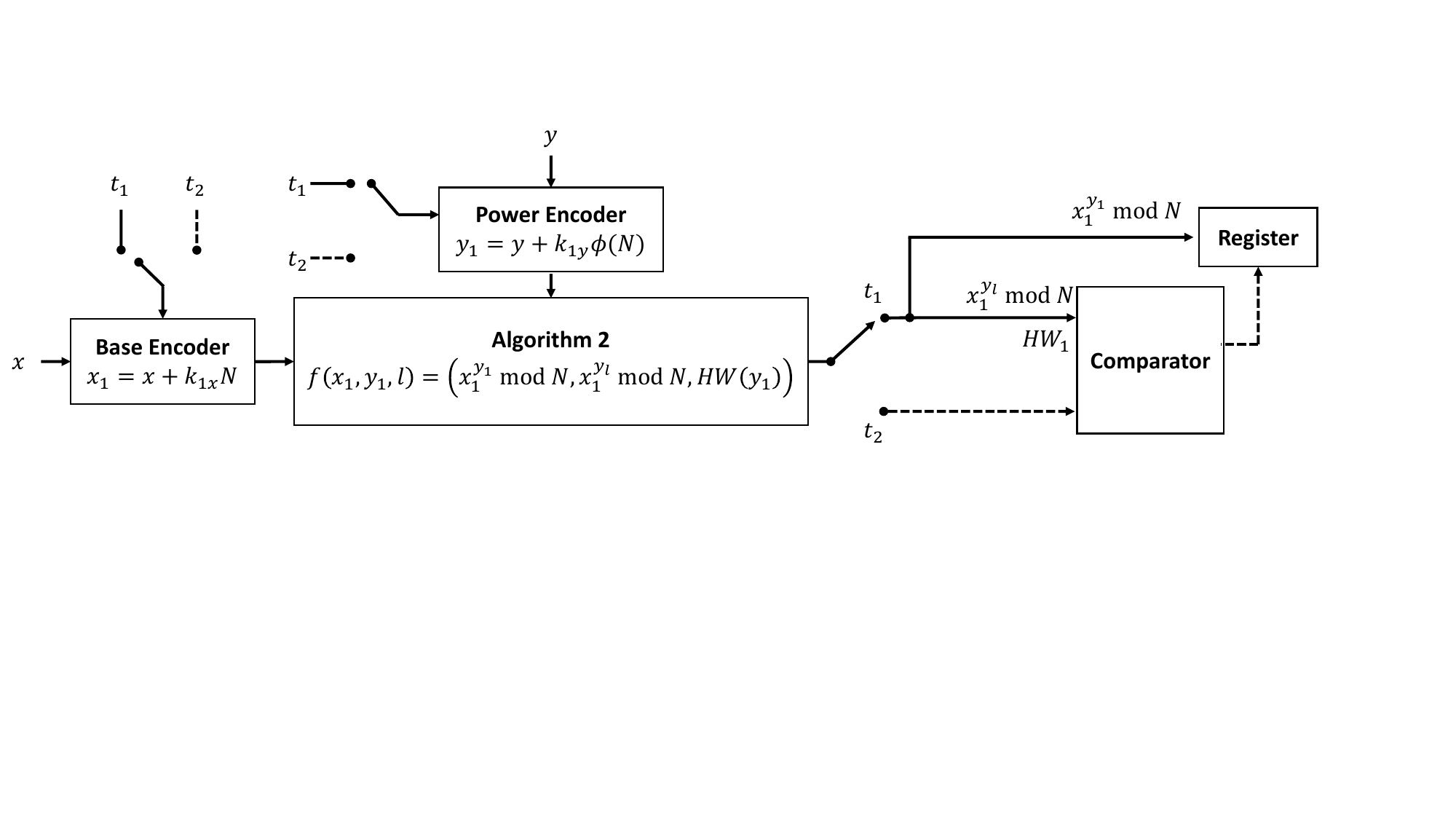}

\caption{Proposed scheme for error detection of modular exponentiation (first
round- main computation).}
\end{figure*}

\begin{figure*}[t]
\centering

\includegraphics[bb=0bp 245bp 960bp 455bp,clip,width=1.8\columnwidth]{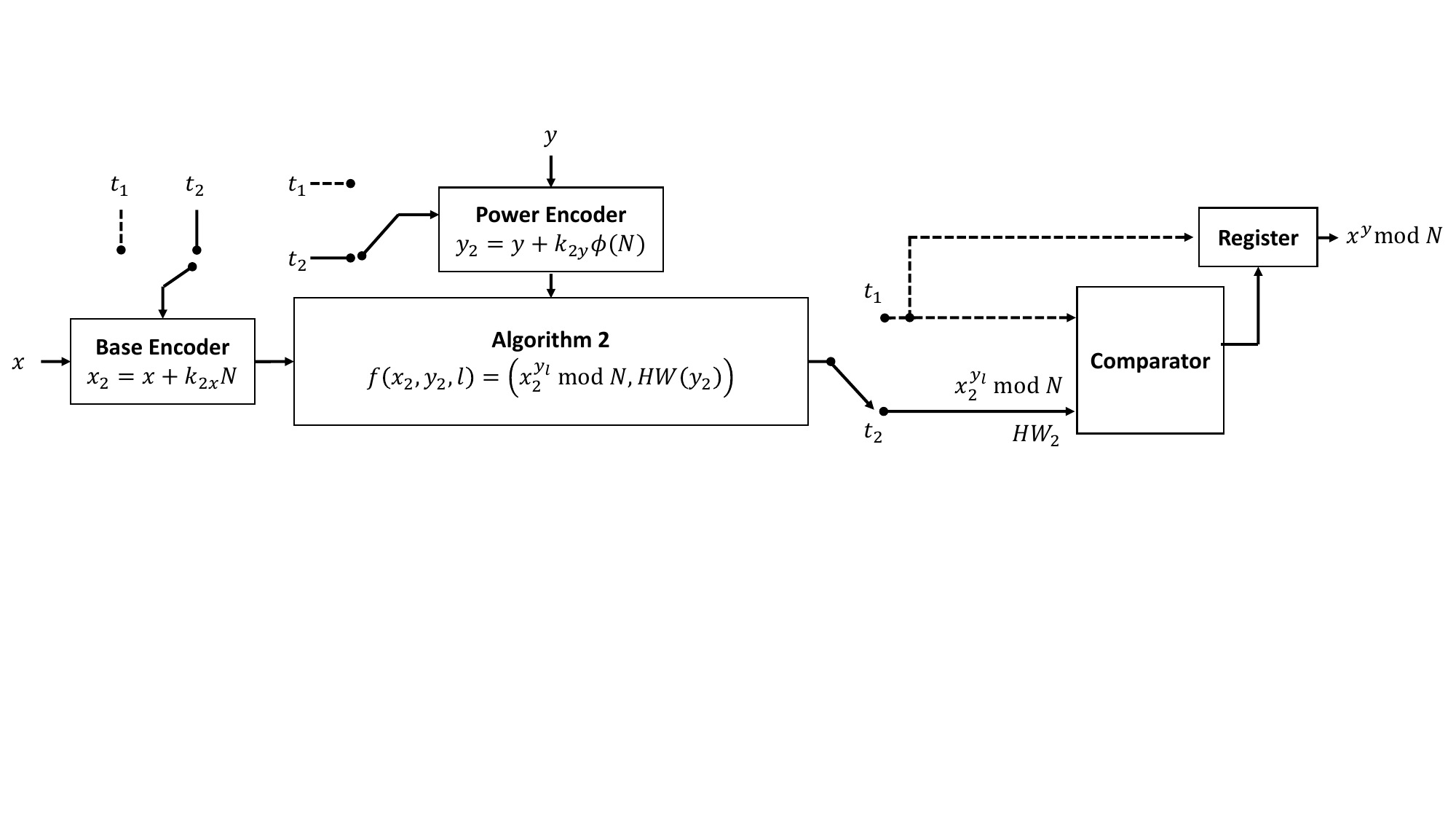}

\caption{Proposed scheme for error detection of modular exponentiation (second
round- partial recomputation).}
\end{figure*}

\subsubsection{Scheme 2: Partial Recomputation}

In this approach, rather than recomputing the $x^{y}\text{ mod}\,N$
at time $t_{2}$, we perform recomputation on a much smaller subset
of the exponent as $y_{partial}=\sum_{i=0}^{l-1}2^{i}y[i]$. We show
that this subset can still effectively detect faults with a high probability.

With more details, at time $t_{1}$ after encoding both the base and
the exponent as $x_{1}=x+k_{1}N$ and $y_{1}=y+k_{2}\phi(N)$, besides
computing $Q_{1}\equiv x_{1}^{y_{1}}\text{ mod}\,N$ another partial
result named $Q_{1,partial}\equiv x_{1}^{y_{1,partial}}\text{ mod}\,N$
is also computed where $y_{1,partial}=\sum_{i=0}^{l-1}2^{i}y_{1}[i]$.
Crucially, $Q_{1,partial}$ does not require any additional computational
steps since it is an intermediate value in the computation of $Q_{1}$.
Therefore, no additional computations is imposed to the algorithm
at $t_{1}$.

At $t_{2}$, after encoding the base and exponent as $x_{2}=x+k_{3}N$
and $y_{2}=y+k_{4}\phi(N)$, instead of calculating $Q_{2}\equiv x_{2}^{y_{2}}\text{ mod}\,N$,
the partial result $Q_{2,partial}\equiv x_{2}^{y_{2,partial}}\text{ mod}\,N$
is computed where $y_{2,partial}=\sum_{i=0}^{l-1}2^{i}y_{2}[i]$.
This adjustment significantly reduces the computational overhead compared
to the first scheme. 

It is worth noting that the effectiveness of this method depends heavily
on the value of $l$. Larger values of $l$ lead to greater error
coverage but also increase the computational overhead. To address
this, we also calculate the Hamming weight of the exponent at both
$t_{1}$ and $t_{2}$ to enhance the error coverage rate, even when
$l$ is relatively small. 

Algorithm 2, Fig. 1 and Fig. 2, demonstrate our design. As a summary,
at $t_{1}$ the values $x_{1}=x+k_{1}N$, $y_{1}=y+k_{2}\phi(N)$,
$y_{1,partial}=\sum_{i=0}^{l-1}2^{i}y_{1}[i]$, $Q_{1}\equiv x_{1}^{y_{1}}\text{ mod}\,N$,
$Q_{1,partial}\equiv x_{1}^{y_{1,partial}}\text{mod}\,N$, and $HM(y_{1}\,\text{mod}\,\phi(N))=HM_{1}$
are computed. Afterward, at $t_{2}$ the values $x_{2}=x+k_{3}N$,
$y_{2}=y+k_{4}\phi(N)$, $y_{2,partial}=\sum_{i=0}^{l-1}2^{i}y_{2}[i]$,
$Q_{2,partial}\equiv x_{2}^{y_{2,partial}}\text{mod}\,N$, and $HM(y_{2}\,\text{mod}\,\phi(N))=HM_{2}$
will be computed. Then after comparing the values $Q_{1,partial}$
with $Q_{2,partial}$, and $HM_{1}$ with $HM_{2}$, $Q_{1}\equiv x_{1}^{y_{1}}\text{mod}\,N$
will be accepted as the output if the corresponding values are equal.

\section{Error Coverage and Simulation Results}

In this section, we conduct several different simulations to evaluate
the error coverage of our design under different fault models to provide
a broad understanding of our method's capabilities.

\subsection{Methodology}

For performing our simulations, we implemented our design on C. To
handle very large numbers, a necessity in real-world applications,
we utilized the GMP library \cite{GMP}, specifically designed for
arithmetic operations involving very large numbers. To comply with
current security constraints, we selected 2048-bit numbers for the
modulus, base, and power. Moreover, the coefficients $k_{i}$ used
for the input encoders, were set to 50 bits in length. Furthermore,
we run each simulation for 1000 iterations.

\subsection{Error coverage}

We have outlined four different fault models namely: total random,
single bit flipping, $k$-bit random flipping, and $k$-bit burst
flipping. In total random model, the inputs are randomly changed to
completely new inputs. In single bit flipping, one bit of the inputs
is chosen randomly and its value is flipped. Similarly, in $k$-bit
random flipping, $k$ bits of inputs are chosen randomly and flipped
meaning if the value was 0 (1) it turns to 1 (0). Finally, in $k$-bit
burst flipping model, after choosing one bit of the inputs randomly,
the next $k$ consecutive bits are flipped. The error coverage results
of our method is demonstrated through Tables I-III. 

It is important to note that even when the output is not faulty in
cases where only the recomputation is altered, our scheme is still
able to detect such changes. This demonstrates the robustness and
reliability of our method, ensuring the integrity of the results even
in scenarios where the output remains correct but the recomputation
process is affected.

\begin{table}[t]
\caption{Simulation Results for ``Total Random'' and ``Single bit Flipping''
models}

\begin{centering}
\centering%
\begin{tabular}{ccccc}
\toprule 
Fault Model & $l$ & \multirow{1}{*}{$x_{1}$} & $y_{1}$ & \multicolumn{1}{c}{$c_{1}^{1}\left\Vert \,c_{2}^{2^{*}}\right\Vert $ $c_{3}^{3}$}\tabularnewline
\midrule
\midrule 
 & $10$ & 99.6\% & 100\% & \tabularnewline
\multirow{2}{*}{Total Random} & $20$ & 100\% & 100\% & \multirow{2}{*}{100\%}\tabularnewline
 & $50$ & 100\% & 100\% & \tabularnewline
 & $128$ & 100\% & 100\% & \tabularnewline
\midrule
\midrule 
 & $10$ & 99.9\% & 98.5\% & \tabularnewline
\multirow{2}{*}{Single bit Flipping} & $20$ & 100\% & 99\% & \multirow{2}{*}{100\%}\tabularnewline
 & $50$ & 100\% & 99.5\% & \tabularnewline
 & $128$ & 100\% & 100\% & \tabularnewline
\bottomrule
\end{tabular}
\par\end{centering}
\begin{raggedright}
\vspace{0.5cm}
\par\end{raggedright}
~

$^{1},^{2}$, $^{3}$ $c_{1}=(x_{1},y_{1}),c_{2}=(x_{2},y_{2}),$
and $c_{3}=(x_{1},x_{2},y_{1},y_{2})$.

$^{^{*}}$In this occasion, although faults occur and detected by
our scheme, they do not introduce output errors.
\end{table}

\begin{table}[t]
\caption{Simulation Results for ``$k$-bit Random Flipping'' model}

\begin{centering}
\centering%
\begin{tabular}{ccccc}
\toprule 
$l$ & number of faults & \multirow{1}{*}{$x_{1}$} & \multirow{1}{*}{$y_{1}$} & \multicolumn{1}{c}{$c_{1}\left\Vert \,c_{2}\right\Vert $ $c_{3}$}\tabularnewline
\midrule
\midrule 
\multirow{6}{*}{$20$} & $3$ & 100\% & 97.9\% & \tabularnewline
 & $5$ & 100\% & 97.1\% & \tabularnewline
 & $15$ & 100\% & 96.9\% & \multirow{2}{*}{100\%}\tabularnewline
 & $25$ & 100\% & 98.2\% & \tabularnewline
 & $75$ & 100\% & 99.1\% & \tabularnewline
 & $128$ & 100\% & 100\% & \tabularnewline
\midrule
\midrule 
\multirow{6}{*}{$50$} & $3$ & 100\% & 98.5\% & \tabularnewline
 & $5$ & 100\% & 99.1\% & \tabularnewline
 & $15$ & 100\% & 98.5\% & \multirow{2}{*}{100\%}\tabularnewline
 & $25$ & 100\% & 98.8\% & \tabularnewline
 & $75$ & 100\% & 100\% & \tabularnewline
 & $128$ & 100\% & 100\% & \tabularnewline
\bottomrule
\end{tabular}
\par\end{centering}
\raggedright{}\vspace{0.5cm}
\end{table}

\begin{table}[t]
\caption{Simulation Results for ``$k$-bit Burst Flipping'' model}

\centering{}\centering%
\begin{tabular}{ccccc}
\toprule 
$l$ & number of faults & \multirow{1}{*}{$x_{1}$} & \multirow{1}{*}{$y_{1}$} & \multicolumn{1}{c}{$c_{1}\left\Vert \,c_{2}\right\Vert $ $c_{3}$}\tabularnewline
\midrule
\midrule 
\multirow{6}{*}{$20$} & $3$ & 100\% & 98.6\% & \tabularnewline
 & $5$ & 100\% & 99.1\% & \tabularnewline
 & $15$ & 100\% & 99.2\% & \multirow{2}{*}{100\%}\tabularnewline
 & $25$ & 100\% & 99.9\% & \tabularnewline
 & $75$ & 100\% & 100\% & \tabularnewline
 & $128$ & 100\% & 100\% & \tabularnewline
\midrule
\midrule 
\multirow{6}{*}{$50$} & $3$ & 100\% & 98.8\% & \tabularnewline
 & $5$ & 100\% & 99.7\% & \tabularnewline
 & $15$ & 100\% & 99.7\% & \multirow{2}{*}{100\%}\tabularnewline
 & $25$ & 100\% & 100\% & \tabularnewline
 & $75$ & 100\% & 100\% & \tabularnewline
 & $128$ & 100\% & 100\% & \tabularnewline
\bottomrule
\end{tabular}
\end{table}

\section{Implementation Results}

To assess the performance of our design, we benchmarked it on both
software and hardware. For software implementation we used ARM Cortex-A72
processor which employs ARMv8 architecture. For hardware we used AMD/Xilinx
Zynq Ultrascale+ and Artix-7 FPGAs. 

\begin{table}[t]
\caption{Total number of clock cycles in 1000 iterations on Cortex-A72 ARM
processor}

\begin{centering}
\centering%
\begin{tabular}{cccc}
\toprule 
$l$ & Unprotected & Our method & Overhead\tabularnewline
\midrule
\midrule 
$10$ & \multirow{5}{*}{39,590,585,862} & 40,362,590,552 & 1.95\%\tabularnewline
$20$ &  & 40,594,147,915 & 2.53\%\tabularnewline
$50$ &  & 41,226,054,689 & 4.13\%\tabularnewline
$128$ &  & 42,625,914,315 & 7.66\%\tabularnewline
$256$ &  & 45,045,633,051 & 13.77\%\tabularnewline
\bottomrule
\end{tabular}
\par\end{centering}
\raggedright{}\vspace{0.3cm}
\end{table}

\subsection{Software Implementation}

To measure the computational overhead of our design, we implemented
it on a Raspberry Pi-4 device. We utilized the C programming language
for implementation and leveraged the GMP library for handling large
numbers, making it suitable for real-world applications. To calculate
the total number of clock cycles incurred by our implementation, we
employed the Performance Application Programming Interface (PAPI)
\cite{PAPI}. Our choice of compiler was clang Version 11. All our
simulation and implementation codes are available on github \footnote{\href{https://github.com/SaeedAghapour/Fault-detection-for-modular-exponentiation}{https://github.com/SaeedAghapour/Fault-detection-for-modular-exponentiation}}.
The results of our Cortex-A72 implementation are presented in Table
IV.

\begin{table}[t]
\caption{Implementation result on AMD/Xilinx Zynq Ultrascale+}

\begin{centering}
\centering%
\begin{tabular}{llccc}
\toprule 
\multirow{2}{*}{Platform} & \multirow{2}{*}{} & \multicolumn{3}{c}{Zynq Ultrascale+}\tabularnewline
 &  & \multicolumn{3}{c}{xczu4ev-sfvc784-2-i}\tabularnewline
\midrule
\multirow{1}{*}{Scheme} &  & Unprotected & Our Method & Overhead\tabularnewline
\midrule
\midrule 
\multirow{3}{*}{Area} & LUT & 12938 & 13039 & 0.78\%\tabularnewline
 & FF & 13469 & 13600 & 0.97\%\tabularnewline
 & DSP & 10 & 10 & -\tabularnewline
\midrule
\multirow{2}{*}{Timing} & Latency (CCs) & 1078 & 1081 & \multirow{2}{*}{0.27\%}\tabularnewline
 & Total Time (ns) & 10780 & 10810 & \tabularnewline
\midrule
\multicolumn{2}{l}{Power $@$100 MHz (W)} & 0.536 & 0.538 & 0.37\%\tabularnewline
\multicolumn{2}{l}{Energy (nJ)} & 5778 & 5816 & 0.65\%\tabularnewline
\bottomrule
\end{tabular}
\par\end{centering}
\raggedright{}\vspace{0.5cm}
\end{table}

\begin{table}[t]
\caption{Implementation result on AMD/Xilinx Artix7}

\begin{centering}
\centering%
\begin{tabular}{llccc}
\toprule 
\multirow{2}{*}{Platform} & \multirow{2}{*}{} & \multicolumn{3}{c}{Artix7}\tabularnewline
 &  & \multicolumn{3}{c}{xa7a100tcsg324-2l}\tabularnewline
\midrule
\multirow{1}{*}{Scheme} &  & Unprotected & Our Method & Overhead\tabularnewline
\midrule
\midrule 
\multirow{3}{*}{Area} & LUT & 12874 & 12907 & 0.25\%\tabularnewline
 & FF & 13633 & 13731 & 0.72\%\tabularnewline
 & DSP & 20 & 20 & -\tabularnewline
\midrule
\multirow{2}{*}{Timing} & Latency (CCs) & 1160 & 1161 & \multirow{2}{*}{0.08\%}\tabularnewline
 & Total Time (ns) & 11600 & 11610 & \tabularnewline
\midrule
\multicolumn{2}{l}{Power $@$100 MHz (W)} & 0.317 & 0.317 & -\tabularnewline
\multicolumn{2}{l}{Energy (nJ)} & 3677 & 3680 & 0.08\%\tabularnewline
\bottomrule
\end{tabular}
\par\end{centering}
\raggedright{}\vspace{0.5cm}
\end{table}

\subsection{Hardware Implementation}

For hardware implementation, we used Xilinx Vivado tool to analyze
area, delay, and power of our design on Zynq Ultrascale+ and Artix7
FPGA families. Both the unprotected and our design utilized identical
settings. We set the clock on 100 MHz and choose $l$ to be 12\% of
the size of the exponent. Tables V and VI presents the result of our
implementation on the Zynq Ultrascale+ and Artix7 FPGAs, respectively.

By combining the implementation and simulation results, we can conclude
that by selecting $l=128$ (6.25\% of $y$'s length), we could achieve
a very high error coverage rate (close to 100\%), while imposing only
a modest computational and area overhead to the design.

\section{Discussions}

In this section, we delve into the practicality of our scheme by highlighting
its relevance to well-known cryptographic applications that heavily
rely on modular exponentiation. In general, a majority of classical
public key cryptography schemes such as Diffie-Hellman key exchange,
RSA cryptosystems, ElGamal encryption, Shamir's secret sharing, and
some PQC schemes, such as KAZ \cite{KAZ} are based on modular exponentiation.

For example, in the Diffie-Hellman protocol, both parties engage in
two modular exponentiations. First, Alice and Bob compute $g^{x}\text{ mod }N$
and $g^{y}\text{ mod }N$, respectively. Subsequently, they establish
a shared key by calculating $(g^{y}\text{ mod}\,N)^{x}\text{ mod}\,N$
and $(g^{x}\text{ mod}\,N)^{y}\text{ mod}\,N$. It is important to
note that when $N=P$ is a prime number, $\phi(N)=p-1$, ensuring
that encodings could be done easily.

Moreover, RSA cryptosystem use modular exponentiation in both encryption
and decryption algorithms. In encryption algorithm, the message $m$
is encrypted through $c=m^{e}$ mod $N$ and the decryption is performed
through $c^{d}$ mod $N=m$. Moreover, since every entity has a key
pair of $(e_{A},d_{A})$ where $e_{A}d_{A}=1\,\text{mod}\,\phi(N)$,
they can perform the encoding because they can obtain a multiplication
of $\phi(N)$ through $e_{A}d_{A}-1=k_{A}\phi(N)$.

\section{Conclusion}

Fault and error detection play a pivotal role in ensuring the integrity
of results within any algorithms. In this paper, we have introduced
a new fault detection approach specifically designed for modular exponentiation,
a critical component in various cryptographic systems such as RSA
and Diffie-Hellman protocols as well as a subset of PQC schemes. What
sets our approach apart is its ability to achieve remarkably high
error detection rates while adding only a minimal computational burden
to the underlying algorithm. Through comprehensive simulations and
real-world implementations on Cortex-A72 ARM processor and two different
FPGAs, we have demonstrated that our method, with a mere 7.66\% increase
in computational cost and less than 1\% in area, can provide high
error coverage. This low overhead underscores the applicability of
our scheme, particularly in resource-constrained embedded devices.

\end{document}